\documentclass[conference,a4paper]{APSIPA2021}
\usepackage{amsmath}
\usepackage{graphicx}
\usepackage{multirow}
\usepackage{threeparttable}

\usepackage{times}
\usepackage{soul}
\usepackage{url}
\usepackage[utf8]{inputenc}
\usepackage{caption}
\usepackage{booktabs}
\usepackage{amsthm}
\usepackage{booktabs}
\usepackage{algorithm}
\usepackage[printonlyused]{acronym}
\usepackage[switch]{lineno}
\usepackage{amsfonts}
\usepackage{amssymb}
\usepackage{xcolor}
\usepackage{setspace}
\usepackage{subcaption}
\usepackage{algorithmic}
\usepackage{hyperref}
\usepackage{cleveref}

\usepackage{geometry}
\usepackage{fancyhdr}

\definecolor{highlightgreen}{HTML}{009901}
\definecolor{highlightred}{HTML}{FD6864}

\acrodef{tta}[TTA]{text-to-audio}
\acrodef{ar}[AR]{Autoregressive}
\acrodef{nar}[NAR]{Non-Autoregressive}
\acrodef{ldm}[LDM]{Latent Diffusion Model}
\acrodef{vae}[VAE]{variational auto-encoder}
\acrodef{method}[SRC-gAudio]{Sample-Rate Controlled Audio Generation}

\acrodef{fd}[FD]{Fréchet distance}
\acrodef{fad}[FAD]{Fréchet audio distance}
\acrodef{is}[IS]{Inception Score}
\acrodef{kl}[KL]{Kullback–Leibler}
\acrodef{clap}[CLAP]{contrastive language-audio pretraining}

\begin{document}

\title{SRC-gAudio: Sampling-Rate-Controlled Audio Generation}

\author{
\authorblockN{
Chenxing Li\authorrefmark{1}, 
Manjie Xu\authorrefmark{1} and
Dong Yu\authorrefmark{2}  \\
}

\authorblockA{
\authorrefmark{1}
Tencent AI Lab, Beijing, China
}

\authorblockA{
\authorrefmark{2}
Tencent AI Lab, Bellevue, WA, USA
}
E-mail: lichenxing007@gmail.com
}

\maketitle
\pagestyle{fancy}

\begin{abstract}
  We introduce SRC-gAudio, a novel audio generation model designed to facilitate text-to-audio generation across a wide range of sampling rates within a single model architecture. SRC-gAudio incorporates the sampling rate as part of the generation condition to guide the diffusion-based audio generation process. Our model enables the generation of audio at multiple sampling rates with a single unified model. Furthermore, we explore the potential benefits of large-scale, low-sampling-rate data in enhancing the generation quality of high-sampling-rate audio. Through extensive experiments, we demonstrate that SRC-gAudio effectively generates audio under controlled sampling rates. Additionally, our results indicate that pre-training on low-sampling-rate data can lead to significant improvements in audio quality across various metrics.
\end{abstract}

\section{Introduction}

\begin{figure*}[t!]
    \centering
    \small
    \includegraphics[width=\linewidth]{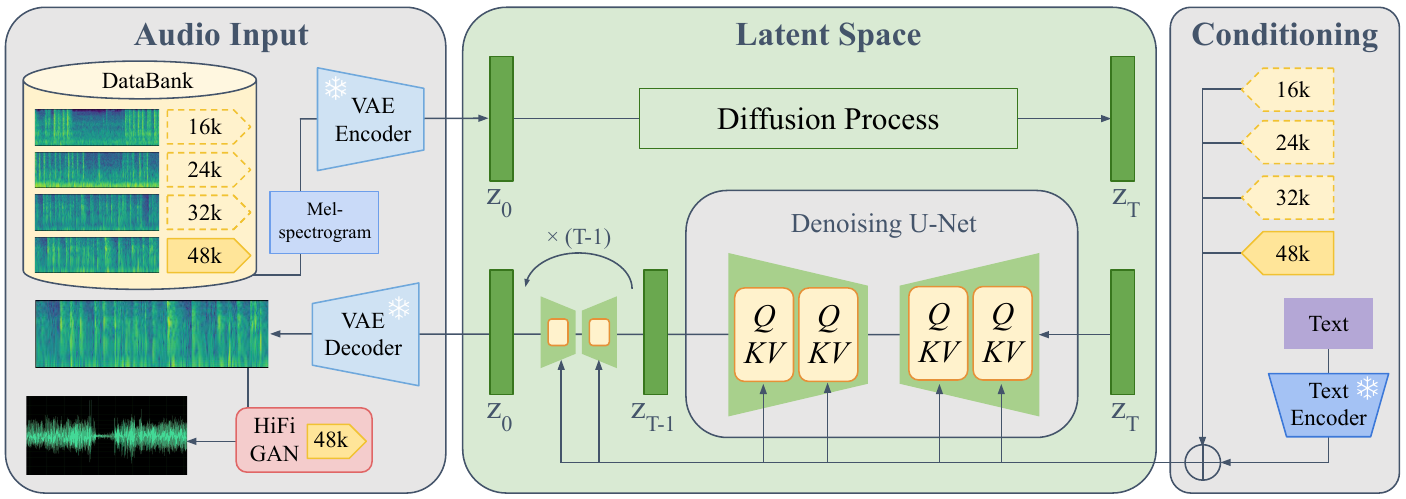}
    \caption{\textbf{The overview of the proposed \ac{method}.}}
    \label{fig:env}
\end{figure*}

The advent of \ac{tta} generation has marked a significant milestone in the realm of multimedia content creation, offering a novel paradigm where textual descriptions are used to generate audio content that aligns with the semantic intent of the text. This technology holds immense potential for enhancing the auditory experience in applications such as audio novels and video dubbing, where the congruence between audio and textual content is paramount. Despite its promise, the field of \ac{tta} generation is fraught with challenges, particularly in the fidelity, quality, and versatility of generated audio.

Current methodologies in text-driven audio generation predominantly bifurcate into two streams: \ac{ar} and \ac{nar} approaches. Audiogen~\cite{kreuk2022audiogen, copet2024simple} and Uniaudio~\cite{yang2023uniaudio} are based on the AR transformer-decoder language model, which condition textual inputs to predict discrete audio tokens step-by-step. For \ac{nar} methods, diffusion-based and flow-based techniques have shown promise in generating high-fidelity audio. Diffsound~\cite{yang2023diffsound} applies diffusion probabilistic models to predict mel-spectrogram tokens. Spectrorgam decoder and vocoder gradually transform tokens into waveform. Notably, diffusion models such as AudioLDM~\cite{liu2023audioldm}, AudioLDM2~\cite{liu2023audioldm2}, Tango~\cite{ghosal2023text}, Make-an-audio~\cite{huang2023make}, and Make-an-audio2~\cite{huang2023make2}, leverage latent variable generation coupled with pre-trained \ac{vae}s~\cite{kingma2013auto} and HiFi-GAN~\cite{kong2020hifi} for audio reconstruction, achieving notable successes. Stable Audio \cite{evans2024fast, evans2024long} directly encodes waveform into latent features and estimates noise updated from U-Net-based \cite{ronneberger2015u} to DiT-based \cite{peebles2023scalable} diffusion model. This solution can simplify the generation process and reduce error accumulation. It also supports longer audio generation. Similarly, flow-based methods, exemplified by Audiobox~\cite{vyas2023audiobox}, employ continuous transformations from simpler to complex data distributions, offering an alternative pathway for audio synthesis. Besides, MAGNet~\cite{ziv2024masked} adopts a \ac{nar} transformer to perform generation directly over several streams of audio tokens. The Hybrid-MAGNet fuses \ac{ar} and \ac{nar} models, which generates the beginning of the tokens in an AR manner while the rest of the sequence is being decoded in parallel. However, there is a performance gap between MAGNet and the methods above in the TTA task.

Despite these advancements, both methodologies encounter inherent limitations, particularly in handling diverse sampling rates and maintaining audio quality. (1) AR-based Audiogen~\cite{kreuk2022audiogen} often suffers from spectral inconsistencies and detail loss due to the not completely accurate prediction of audio tokens. Also, the pre-trained sampling rate specified EnCodec~\cite{defossez2022high} limits the sampling rate of audio generated by Audiogen. (2) The current NAR methods have guaranteed generation performance, but they all generate audio at a sampling rate of 16 kHz, which sometimes can not meet the high-resolution requirements. Under the popular diffusion-based pipeline, like AudioLDM2~\cite{liu2023audioldm2}, modeling different sampling rates requires training \ac{vae}, U-Net, and vocoder separately, which will increase the complexity of training and deployment. These limitations not only impact the listening experience but also underscore the need for a more flexible and robust framework capable of addressing the multifaceted challenges of audio generation. 

An intuitive method for generating audio with varying sampling rates involves training a model at a high sampling rate, for instance, 48 kHz, and then down-sampling as per the needs of real-world applications. However, the scarcity of high-quality, high-sampling-rate annotated data often hampers adequate training for high-sampling-rate models. Moreover, under high-resolution conditions, the model may struggle to capture the wider range and increased variability of frequency features. Consequently, such models often underperform in practical scenarios.

This situation underscores the need for a versatile approach to overcome these limitations. Recognizing these challenges, we propose the \ac{method}, a multi-sampling-rate controlled audio generation model that uses the sampling rate as a condition to control generation within a single model. Moreover, \ac{method} leverages the strengths of low-sampling-rate pre-trained models to enhance generation quality at higher sampling rates. In detail, \ac{method} comprises the following components:
\begin{enumerate}
    \item Diffusion-based SRC-gAudio fuses the sampling rate and text prompt as a condition to control audio generation. Through joint training under multi-sampling-rate conditions, \ac{method} shares the pre-trained text encoder, \ac{vae}, and trainable diffusion model. There is no need to train these models separately for different sampling rates, reducing the complexity of the overall system. The HiFi-GAN-based vocoder still needs to be trained separately.
    \item Due to the scarcity of high-resolution data and the increased difficulty of training high-resolution models, we use low-sampling-rate data to help train high-sampling-rate models. We first employ large-scale low-sampling-rate data to train the model and then conduct multi-sampling-rate training based on the pre-trained model. A model trained with low-sampling-rate data can help the model reach a good initial point, and large-scale data can assist the model in achieving better generalization performance.
\end{enumerate}

Our model offers a versatile solution that adapts to various sampling rates without compromising audio fidelity. Through rigorous experimentation, based on objective and subjective evaluations, \ac{method} demonstrates significant improvements, paving the way for a new era in sampling-rate-controlled audio generation technology.

\section{System overview}

\subsection{Conditional LDM-based audio generation}

The \ac{ldm} has emerged as a promising approach for generating high-quality and diverse audio samples in \ac{tta} task~\cite{liu2023audioldm, ghosal2023text, huang2023make}. Given a data point $x_0$ sampled from the real data, \ac{ldm}s focus on the efficient, low-dimensional latent space, with a trained perceptual compression model mapping the input to a hidden continuous feature space $Z$. In \ac{tta} generation, $z \in Z$ is the latent representation of the mel-spectrogram of the audio $x$. We leverage a pre-trained audio \ac{vae} from AudioLDM~\cite{liu2023audioldm} to help compress the mel-spectogram of an audio sample $x$ into the latent space. 

In the forward process, Gaussian noise is gradually added to the input data according to a variance schedule $\beta$:
\begin{equation}
    q(z_t|z_{t-1}) :=  \mathcal{N} \left( z_t; \sqrt{1 - \beta_t} z_{t-1} , \beta_t I \right).
\end{equation}
In reverse diffusion, diffusion models iteratively refine a randomly sampled noise input from $z_{t} \sim \mathcal{N}(0, I)$ to $z_0$. There have been methods that incorporate guidance into the diffusion process in order to ``guide" the generation. The guidance refers to conditioning a latent of the prior data distribution $p(z)$ with a condition $c$, i.e., the class label or an image/text embedding, resulting in $p(z|c)$. In our model, the generation guidance $c$ is included in the reverse process through cross-attention, where $K$ and $V$ embeddings are replaced with the condition embedding. In existing \ac{tta} models like AudioLDM~\cite{liu2023audioldm} and Tango~\cite{ghosal2023text}, the condition is the text embedding of the textual prompt. In the \ac{method}, the model functions by also conditioning the generation process on the desired sampling rate. We concatenate the sampling rate label, denoted as $Sr$, with the text condition $\psi(P)$ to form a new condition $c = concat(\psi(P), Sr)$. To turn a diffusion model $p_\theta$ into a conditional diffusion model, we add conditioning information at each diffusion step:
\begin{equation}
    p_\theta (z_{0:T}|\psi(P), Sr) = p_\theta(z_T) \prod_{t=1}^{T} p_\theta (z_{t-1}|z_t, c).
\end{equation}

With a given text description $P$, to perform sequential denoising, a network $\epsilon_\theta$ is often trained to predict artificial noise, following the objective:
\begin{equation}
\min_\theta \mathbb{E}_{z_0, \epsilon \sim \mathcal{N}(0, I), t \sim \text{Uniform}(1, T)} \lVert \epsilon - \epsilon_\theta(z_t, t, c) \rVert_2^2,
\label{eq:loss1}
\end{equation}
where the $\psi(P)$ is the text embedding of the description. We leverage a pre-trained FLAN-T5 model ~\cite{chung2022scaling} as the text encoder to obtain the text embedding.

We use U-Net~\cite{ronneberger2015u,rombach2022high} with a cross-attention component as the backbone model for noise estimation. In detail, the U-Net model is conditioned on both the time step, sampling rate, and text embedding. We map the time step and sampling rate into a one-dimensional embedding and then concatenate these with text embedding as conditioning information. This approach allows for the generation of high-quality audio across a wide range of resolutions, all within a unified framework. The generation process is guided by these conditioning variables, ensuring that the output is in line with the attributes of the desired sampling rate.

We also leverage classifier-free guidance~\cite{ho2022classifier} to guide the diffusion toward the target audio. To achieve that, let $\varnothing = \psi(\text{" "})$ be the null text embedding, we define $c_\varnothing = concat(\psi(\text{" "}), Sr)$ as the unconditional case, and thus the generation can be defined by:
\begin{equation}
\tilde{\epsilon}_\theta = w \cdot \epsilon_\theta(z_t, t, c) + (1 - w) \cdot \epsilon_\theta(z_t, t, c_\varnothing),
\end{equation}
where $w$ denotes the guidance scale.

\subsection{The pipeline of SRC-gAudio}

The schematic diagram of SRC-gAudio is shown in Figure~\ref{fig:env}, which contains modules including a mel-spectrogram extraction module, an audio \ac{vae}, a text encoder, a \ac{ldm}, and sampling-rate-based vocoders. 

In the training stage, the mel-spectrogram is first extracted from the audio, and then the audio \ac{vae} is used to convert the mel-spectrogram into audio latents. The U-Net is trained based on Equ.(3) under text description and sampling rate control. In the generation stage, the audio latent is first initialized from $\mathcal{N}(0, I)$. Through the diffusion process, under the guidance of the text prompt and sampling rate conditions, the latent of target audio is gradually generated. The decoder of \ac{vae} and HiFi-GAN vocoder restore audio latent to mel-spectrogram and audio waveform step by step.

We mainly train the \ac{ldm} under the control of different sampling rates. We also train vocoders so that they can convert the mel-spectrogram of audios with different sampling rates to audio. 

\subsection{Pre-training on low-sampling-rate data}

Compared with low-sampling-rate data, high-sampling-rate data contains more high-frequency details and greater energy differences among frequencies, which may barricade model convergence. High sampling rate data, such as 32 kHz or 48 kHz, is difficult to obtain. The limited size of high-sampling rate data further leads to insufficient training of high-sampling rate-based generation models. These may lead to problems such as poor generation quality and less similarity between the generated audio and text prompt.

\ac{method} aims to generate audio with different sampling rates. In practice, the amount of data at different sampling rates is different. The amount of data in 16 kHz is much larger than the amount of data in 32 kHz or 48 kHz. We intuitively consider using a large amount of low-sampling-rate data to help generate high-sampling-rate audio. Specifically, we use low-sampling-rate data to pre-train a model with a fixed sampling rate. After the model training is completed, \ac{method} model training is continued based on the pre-trained model.

\section{Experimental setup}

\begin{table}[t]
  \caption{The evaluation results of the \ac{method} in two training paradigms: sampling rate as the generation condition (top) or training separately on different sampling rates (below).}
  \label{tab:1}
  \centering
  \resizebox{\columnwidth}{!}{
  \begin{tabular}{ccccccc}
    \toprule[1pt]
    Models & Sample rate & FD{\color{highlightgreen} $\downarrow$} &IS {\color{highlightred} $\uparrow$}& KL{\color{highlightgreen} $\downarrow$} & FAD{\color{highlightgreen} $\downarrow$} & CLAP{\color{highlightred} $\uparrow$} \\
    \hline
    \hline
    Ground-truth & - &- &11.24 &- &- &0.501 \\
    \hline
    \multicolumn{1}{c}{\multirow{4}{*}{\text{SRC-gAudio}}}
    & 16k &26.63 &7.12 &1.34 &1.93 & 0.505 \\
    & 24k &35.15 &6.77 &1.44 &2.07 & 0.611 \\
    & 32k &45.40 &6.85 &1.52 &2.09 & 0.592 \\
    & 48k &56.79 &4.99 &1.65 &5.90 & 0.569 \\
    \hline
    gAudio & 16k &28.76	&7.49 &1.74 &2.70 & 0.402\\
    gAudio & 24k &35.91	&6.70 &1.60	&2.66 & 0.427 \\
    gAudio & 32k & 44.57 &6.08 &1.81 &3.53 & 0.540 \\
    gAudio & 48k & 56.87 &4.88 &1.80 &9.95 & 0.549 \\
    \bottomrule[1pt]
  \end{tabular}
  }
\end{table}

\subsection{Dataset}

In this study, we primarily utilize the Audiocaps dataset~\cite{kim2019audiocaps} as the foundation for training. We obtain the audio data at a 48 kHz sample rate, which consists of 41,597 files\footnote{A few audio clips are excluded due to broken download links.}. The dataset comprises 40259 clips (120 hours in total) in the training set, 381 clips (1 hour in total) in the validation set, and 957 clips (3 hours in total) in the test set. During the model pre-training phase, we collect data from WavCaps~\cite{mei2023wavcaps}, VGGSound~\cite{chen2020vggsound}, and ESC~\cite{piczak2015esc}, amounting to approximately 4,000 hours of audio with paired captions. The training dataset for the 16k, 24k, and 32khz-based HiFi-GAN vocoder is the same as the pre-train dataset. For the 48khz-based vocoder, the train data is selected from freesound\footnote{https://freesound.org/}, which is extracted from WavCaps. In the test set, the caption for each clip is consistent with Tango~\cite{ghosal2023text} and AudioLDM~\cite{liu2023audioldm}. This consistency ensures a fair comparison with their respective works.

\begin{table*}[t]
  \caption{The evaluation results of the pre-train-based \ac{method} and the comparisons with other baseline methods. pre-AC16k+ft-AC and pre-Full16k+ft-AC indicate \ac{method} first pra-trains on Audiocaps or full pre-training data with fixed 16 kHz sampling rate and then fine-tunes on Audiocaps with sampling-rate condition, respectively. For baselines, 48 kHz-based gAudio-FT first pre-trains on 48 kHz upsampled full pre-training data and fine-tunes on 48 kHz-based Audiocaps. FT means the model is fine-tuned on the Audiocaps dataset. AC and AS refer to the Audiocaps and Audioset~\cite{gemmeke2017audio} dataset.}
  \label{tab:2}
  \centering
  \resizebox{2.0\columnwidth}{!}{
  \begin{tabular}{cccccccccccc}
    \toprule[1pt]
    \multirow{2}{*}{Models} & \multirow{2}{*}{Dataset} & \multirow{2}{*}{Params} & \multirow{2}{*}{Sample rate} & \multicolumn{5}{c}{Objective metrics} & \multicolumn{3}{c}{Subjective metrics} \\
    &&& &FD {\color{highlightgreen} $\downarrow$} &IS {\color{highlightred} $\uparrow$} & KL{\color{highlightgreen} $\downarrow$} & FAD{\color{highlightgreen} $\downarrow$} & CLAP{\color{highlightred} $\uparrow$} &OGL {\color{highlightred} $\uparrow$} &RL {\color{highlightred} $\uparrow$} &AQ {\color{highlightred} $\uparrow$}\\
    \hline
    \hline
    Ground-truth &- &- & - &-  &11.24 &- &- &0.501 &93.72 &93.64 &94.45\\
    \hline
    \multicolumn{1}{c}{\multirow{4}{*}{\text{SRC-gAudio-FT}}}
    &\multicolumn{1}{c}{\multirow{4}{*}{\text{pre-AC16k+ft-AC}}}
    &\multicolumn{1}{c}{\multirow{4}{*}{\text{561 M}}}
    & 16k &23.45	&8.14 	&1.31	&1.87 & 0.534 &92.06 &92.33 &93.57\\
    & & & 24k &30.03	&7.85 	&1.32	&2.44 & 0.627 &92.16 &91.62 &94.02\\
    & & & 32k &38.70	&7.75	&1.43	&2.24 & 0.618 &92.93 &92.43 &93.09\\
    & & & 48k &53.29	&5.37	&1.58	&4.64 & 0.578 &92.34 &91.26 &93.02\\
    \hline
    \multicolumn{1}{c}{\multirow{4}{*}{\text{SRC-gAudio-FT}}}
    &\multicolumn{1}{c}{\multirow{4}{*}{\text{pre-Full16k+ft-AC}}}
    &\multicolumn{1}{c}{\multirow{4}{*}{\text{561 M}}}
    & 16k &20.63	&8.91	&1.21	&2.10 &0.529 &92.81 &92.56 &93.56\\
    & & & 24k &26.83	&8.50	&1.30	&3.00 &0.633 &93.58 &92.74 &94.36\\
    & & & 32k &35.80	&9.15	&1.38	&2.62 &0.625 &92.59 &92.21 &94\\
    & & & 48k &49.67	&5.83	&1.51	&4.51 &0.568 &91.26	&93.34 &92.75\\
    \hline
    gAudio-FT &pre-Full48k+ft-AC &561 M & 48k &51.96 &3.45 &2.29 &4.18 &0.533 &89.14 &90.64 &90.58\\
    \hline
    Tango  &AC &\multicolumn{1}{c}{\multirow{2}{*}{\text{866 M}}}  & 16k &25.50 &7.04 &1.35 &1.82 & 0.491 &92.2 &91.78 &94.3\\
    Tango-Full-FT & AS+AC+6 others& & 16k &18.43 &8.01 &1.16 &2.77 & 0.554 &92.89 &92.46 &94.05\\
    AudioLDM-L-Full &AS+AC+2 others &\multicolumn{1}{c}{\multirow{2}{*}{\text{739 M}}}  & 16k &30.90 &7.59 &1.64 &4.61 &0.427 &90.27 &89.04 &91.17 \\
    AudioLDM-L-Full-FT &AS+AC+2 others & & 16k & 23.31 &8.13 &1.59 &1.96 &- &- &- &-\\
    AudioLDM2-Full 
    & AS+AC+6 others &346 M & 16k &25.75 &8.28 &1.58 &3.39 &0.435 &92.1	&89.84 &92.27\\
    AudioLDM2-48k &- &262 M & 48k &62.75	&5.91	&2.19	&4.75 & 0.526 &89.65 &88.54 &91.39\\
    \bottomrule[1pt]
  \end{tabular}
  }
\end{table*}

\subsection{Model architecture}
We use the pre-trained audio \ac{vae} from AudioLDM~\cite{liu2023audioldm} and the frozen FLAN-T5-LARGE~\cite{chung2022scaling} text encoder during training. For mel-spectogram extraction, we perform feature extraction in the (fftSize, hopSize, melDim) format for 16k, 24k, 32k, and 48khz-based configurations as follows: (1024, 160, 64),  $(2048, 240, 64)$, $(2048, 320, 64)$, and $(2048, 480, 64)$, respectively. The 16k, 24k, 32k, and 48khz-based HiFi-GAN vocoders adopt the same structure as the vocoder in AudioLDM~\cite{liu2023audioldm} but are trained separately with different sample rates.

We adopt the U-Net backbone of StableDiffusion~\cite{rombach2022high} as the basic architecture of LDM of SRC-gAudio. The U-Net backbone we use has four encoder blocks, a middle block, and four decoder blocks. The channel dimensions of encoder blocks are $[320, 640, 640, 1280]$. The channel dimensions of decoder blocks are the reverse of encoder blocks. We add a cross-attention block in the last three encoder blocks and the first three decoder blocks, in which the number of heads is $[5, 10, 10, 20]$. The LDM of SRC-gAudio occupies 561M parameters.

We train the model for 40 epochs and report results for the checkpoint with the best validation loss. In sampling, we employ the DDIM~\cite{song2020score} sampler with 200 sampling steps. For classifier-free guidance, a guidance scale $w$ of 3.0 is used.

\subsection{Evaluation metric}

\subsubsection{Objective metric}
We employed various metrics, including \ac{fd}, \ac{kl}, \ac{is}, \ac{fad}, and \ac{clap}\footnote{630k-best checkpoint from https://github.com/LAION-AI/CLAP} \cite{wu2023large} to evaluate its performance. \ac{fd} and \ac{fad} are used to measure the similarity distance between the generated audio and the ground truth audio, while \ac{kl} calculates the Kullback–Leible divergence between them. Lower values for these metrics indicate better performance. On the other hand, \ac{is} is employed to measure the diversity and quality of the generated audio. \ac{clap} evaluates the similarity between the generated audio and the text description. 

\subsubsection{Subjective metric}

For subjective evaluation, we recruit 8 human participants to conduct a rating process. Following a similar approach to \cite{yang2023diffsound}, the generated samples are assessed based on the overall generation quality (OGL), relevance to the input text (REL), and audio quality (AQ) using a scale of 1 to 100. Specifically, OGL examines the overall impact of the generated sound effect, encompassing factors such as semantic consistency between audio and text, and audio quality. REL measures the completeness and sequential consistency between the generated audio and the text prompt. AQ evaluates the quality of the generated audio, including aspects like audio clarity and intelligibility. We randomly select 100 test audio samples from the AudioCaps test set. Each participant is asked to evaluate each audio sample.

\subsection{Baselines}

To validate the effectiveness of the proposed SRC-gAudio approach, we initially train the generation model, named gAudio, using data with fixed sampling rates. The model structure and training strategy of gAudio mirror those of SRC-gAudio. However, without sampling rate control, gAudio can only generate audio at its pre-determined sampling rate.

For model comparisons, we employ diffusion-based generative models, including Tango~\cite{ghosal2023text}, AudioLDM~\cite{liu2023audioldm}, and AudioLDM2~\cite{liu2023audioldm2}. Currently, these generation models mainly generate audio with a 16 kHz sampling rate, with only AudioLDM2 offering a 48 kHz version. We adopt the results from AudioLDM, denoted as AudioLDM-L-Full-FT. For Tango and Tango-Full-FT\footnote{https://github.com/declare-lab/tango}, AudioLDM-L-Full\footnote{https://github.com/haoheliu/AudioLDM}, AudioLDM2-Full, and AudioLDM2-48k\footnote{https://github.com/haoheliu/AudioLDM2}, we evaluate the results using the models and code released by the authors. The baseline results may differ from those in the corresponding papers, which could be attributed to the randomness of the generation process. Furthermore, to assess only the effectiveness of the generation model, we do not include clap filtering \cite{liu2023audioldm2}, which also accounts for the discrepancies between the baseline results and those reported in the papers.

\section{Experiments results}

\subsection{Experimental results of \ac{method}}

The evaluation results of the \ac{method} are presented in Table~\ref{tab:1}. We train our model under two paradigms: using the sampling rate as the generation condition to do joint training (top portion of the table) and training separately for different sampling rates (bottom portion of the table). Generally, lower sampling rates yield better generation results in evaluation metrics, indicating that audios with higher sampling rates are more challenging to learn. The results demonstrate that the \ac{method} can achieve competitive generation outcomes when trained collectively across various sampling rates, as compared to training a single model specifically for each sampling rate. A key observation is that in the case of high sampling rates, such as 32 kHz and 48 kHz, the joint-training \ac{method} outperforms the model trained specifically for the sampling rate in most evaluation metrics. These results suggest that training on lower sampling rates, which are considered easier to learn, can help the model fit better in more difficult generation tasks.

\subsection{Experimental results on pre-training}
The outcomes of the \ac{method} with pre-training are displayed in Table~\ref{tab:2}. Compared Table~\ref{tab:2} with Table~\ref{tab:1}, it's evident that pre-training on data with a low sampling rate significantly enhances the generation of audio for both sampling rates. Pre-training was conducted on the identically distributed dataset (Audiocaps-16k) and large-scale pre-training dataset (Full16k) followed by fine-tuning. Based on objective evaluation, both tests reveal that a \ac{method} pre-trained on 16 kHz data yields superior generation results, particularly at higher sampling rates such as 32 kHz and 48 kHz. 

After pre-training on the upsampled large-scale pre-training dataset and fine-tuned on Audiocaps, 48 kHz-based gAudio-FT still performes worse than the corresponding \ac{method}-FT. This illustrates the difficulty of training high-sampling rate models using low-quality data and further demonstrates the effectiveness of our proposed pipeline.

Additionally, we contrast the generation outcomes of \ac{method} with other cutting-edge baseline models such as Tango, AudioLDM, and AudioLDM2. By making the sampling rate an essential aspect of the generation condition, from both the objective and subjective metrics, our model delivers competitive results across different sampling rates in comparison to state-of-the-art baselines. It is observed that \ac{method} surpasses AudioLDM2-48k in generating high-sampling-rate audio. Even in the absence of pre-training (as depicted in Table~\ref{tab:1}), \ac{method} excels in generating 48 kHz audio. We emphasize that training on low sampling-rate data can be beneficial.

\section{Conclusion}

In this study, we present \ac{method}, an audio generation model designed to accommodate multiple sampling rates within a single, unified model framework. \ac{method} is capable of producing audio outputs at various sampling rates, conditioning on the specific generation configurations provided. By incorporating the sampling rate and text prompt as conditions, \ac{method} is jointly trained under multi-sampling-rate conditions. Furthermore, \ac{method} leverages the strengths of low-sampling-rate pre-trained models to enhance the generation quality at higher sampling rates. Our approach offers a scalable solution that adapts to diverse sampling rates without compromising audio fidelity. Rigorous experimentation has demonstrated significant improvements in TTA generation, paving the way for advancements in sampling-rate-controlled audio generation technology.


\bibliographystyle{IEEEbib}
\bibliography{mybib}

\begin{thebibliography}{10}

\bibitem{kreuk2022audiogen}
Felix Kreuk, Gabriel Synnaeve, Adam Polyak, Uriel Singer, Alexandre
  D{\'e}fossez, Jade Copet, Devi Parikh, Yaniv Taigman, and Yossi Adi,
\newblock ``Audiogen: Textually guided audio generation,''
\newblock in {\em The Eleventh International Conference on Learning
  Representations}, 2022.

\bibitem{copet2024simple}
Jade Copet, Felix Kreuk, Itai Gat, Tal Remez, David Kant, Gabriel Synnaeve,
  Yossi Adi, and Alexandre D{\'e}fossez,
\newblock ``Simple and controllable music generation,''
\newblock {\em Advances in Neural Information Processing Systems}, vol. 36,
  2024.

\bibitem{yang2023uniaudio}
Dongchao Yang, Jinchuan Tian, Xu~Tan, Rongjie Huang, Songxiang Liu, Xuankai
  Chang, Jiatong Shi, Sheng Zhao, Jiang Bian, Xixin Wu, et~al.,
\newblock ``Uniaudio: An audio foundation model toward universal audio
  generation,''
\newblock {\em arXiv preprint arXiv:2310.00704}, 2023.

\bibitem{yang2023diffsound}
Dongchao Yang, Jianwei Yu, Helin Wang, Wen Wang, Chao Weng, Yuexian Zou, and
  Dong Yu,
\newblock ``Diffsound: Discrete diffusion model for text-to-sound generation,''
\newblock {\em IEEE/ACM Transactions on Audio, Speech, and Language
  Processing}, 2023.

\bibitem{liu2023audioldm}
Haohe Liu, Zehua Chen, Yi~Yuan, Xinhao Mei, Xubo Liu, Danilo Mandic, Wenwu
  Wang, and Mark~D Plumbley,
\newblock ``{AudioLDM}: Text-to-audio generation with latent diffusion
  models,''
\newblock {\em Proceedings of the International Conference on Machine
  Learning}, 2023.

\bibitem{liu2023audioldm2}
Haohe Liu, Qiao Tian, Yi~Yuan, Xubo Liu, Xinhao Mei, Qiuqiang Kong, Yuping
  Wang, Wenwu Wang, Yuxuan Wang, and Mark~D Plumbley,
\newblock ``Audioldm 2: Learning holistic audio generation with self-supervised
  pretraining,''
\newblock {\em arXiv preprint arXiv:2308.05734}, 2023.

\bibitem{ghosal2023text}
Deepanway Ghosal, Navonil Majumder, Ambuj Mehrish, and Soujanya Poria,
\newblock ``Text-to-audio generation using instruction guided latent diffusion
  model,''
\newblock in {\em Proceedings of the 31st ACM International Conference on
  Multimedia}, 2023, pp. 3590--3598.

\bibitem{huang2023make}
Rongjie Huang, Jiawei Huang, Dongchao Yang, Yi~Ren, Luping Liu, Mingze Li,
  Zhenhui Ye, Jinglin Liu, Xiang Yin, and Zhou Zhao,
\newblock ``Make-an-audio: Text-to-audio generation with prompt-enhanced
  diffusion models,''
\newblock {\em arXiv preprint arXiv:2301.12661}, 2023.

\bibitem{huang2023make2}
Jiawei Huang, Yi~Ren, Rongjie Huang, Dongchao Yang, Zhenhui Ye, Chen Zhang,
  Jinglin Liu, Xiang Yin, Zejun Ma, and Zhou Zhao,
\newblock ``Make-an-audio 2: Temporal-enhanced text-to-audio generation,''
\newblock {\em arXiv preprint arXiv:2305.18474}, 2023.

\bibitem{kingma2013auto}
Diederik~P Kingma and Max Welling,
\newblock ``Auto-encoding variational bayes,''
\newblock {\em arXiv preprint arXiv:1312.6114}, 2013.

\bibitem{kong2020hifi}
Jungil Kong, Jaehyeon Kim, and Jaekyoung Bae,
\newblock ``Hifi-gan: Generative adversarial networks for efficient and high
  fidelity speech synthesis,''
\newblock {\em Advances in Neural Information Processing Systems}, vol. 33, pp.
  17022--17033, 2020.

\bibitem{evans2024fast}
Zach Evans, CJ~Carr, Josiah Taylor, Scott~H Hawley, and Jordi Pons,
\newblock ``Fast timing-conditioned latent audio diffusion,''
\newblock {\em arXiv preprint arXiv:2402.04825}, 2024.

\bibitem{evans2024long}
Zach Evans, Julian~D Parker, CJ~Carr, Zack Zukowski, Josiah Taylor, and Jordi
  Pons,
\newblock ``Long-form music generation with latent diffusion,''
\newblock {\em arXiv preprint arXiv:2404.10301}, 2024.

\bibitem{ronneberger2015u}
Olaf Ronneberger, Philipp Fischer, and Thomas Brox,
\newblock ``U-net: Convolutional networks for biomedical image segmentation,''
\newblock in {\em Medical Image Computing and Computer-Assisted
  Intervention--MICCAI 2015: 18th International Conference, Munich, Germany,
  October 5-9, 2015, Proceedings, Part III 18}. Springer, 2015, pp. 234--241.

\bibitem{peebles2023scalable}
William Peebles and Saining Xie,
\newblock ``Scalable diffusion models with transformers,''
\newblock in {\em Proceedings of the IEEE/CVF International Conference on
  Computer Vision}, 2023, pp. 4195--4205.

\bibitem{vyas2023audiobox}
Apoorv Vyas, Bowen Shi, Matthew Le, Andros Tjandra, Yi-Chiao Wu, Baishan Guo,
  Jiemin Zhang, Xinyue Zhang, Robert Adkins, William Ngan, et~al.,
\newblock ``Audiobox: Unified audio generation with natural language prompts,''
\newblock {\em arXiv preprint arXiv:2312.15821}, 2023.

\bibitem{ziv2024masked}
Alon Ziv, Itai Gat, Gael~Le Lan, Tal Remez, Felix Kreuk, Alexandre Défossez,
  Jade Copet, Gabriel Synnaeve, and Yossi Adi,
\newblock ``Masked audio generation using a single non-autoregressive
  transformer,'' 2024.

\bibitem{defossez2022high}
Alexandre D{\'e}fossez, Jade Copet, Gabriel Synnaeve, and Yossi Adi,
\newblock ``High fidelity neural audio compression,''
\newblock {\em arXiv preprint arXiv:2210.13438}, 2022.

\bibitem{chung2022scaling}
Hyung~Won Chung, Le~Hou, Shayne Longpre, Barret Zoph, Yi~Tay, William Fedus,
  Yunxuan Li, Xuezhi Wang, Mostafa Dehghani, Siddhartha Brahma, Albert Webson,
  Shixiang~Shane Gu, Zhuyun Dai, Mirac Suzgun, Xinyun Chen, Aakanksha
  Chowdhery, Alex Castro-Ros, Marie Pellat, Kevin Robinson, Dasha Valter,
  Sharan Narang, Gaurav Mishra, Adams Yu, Vincent Zhao, Yanping Huang, Andrew
  Dai, Hongkun Yu, Slav Petrov, Ed~H. Chi, Jeff Dean, Jacob Devlin, Adam
  Roberts, Denny Zhou, Quoc~V. Le, and Jason Wei,
\newblock ``Scaling instruction-finetuned language models,'' 2022.

\bibitem{rombach2022high}
Robin Rombach, Andreas Blattmann, Dominik Lorenz, Patrick Esser, and Bj{\"o}rn
  Ommer,
\newblock ``High-resolution image synthesis with latent diffusion models,''
\newblock in {\em Proceedings of the IEEE/CVF conference on computer vision and
  pattern recognition}, 2022, pp. 10684--10695.

\bibitem{ho2022classifier}
Jonathan Ho and Tim Salimans,
\newblock ``Classifier-free diffusion guidance,''
\newblock {\em arXiv preprint arXiv:2207.12598}, 2022.

\bibitem{kim2019audiocaps}
Chris~Dongjoo Kim, Byeongchang Kim, Hyunmin Lee, and Gunhee Kim,
\newblock ``Audiocaps: Generating captions for audios in the wild,''
\newblock in {\em Proceedings of the 2019 Conference of the North American
  Chapter of the Association for Computational Linguistics: Human Language
  Technologies, Volume 1 (Long and Short Papers)}, 2019, pp. 119--132.

\bibitem{mei2023wavcaps}
Xinhao Mei, Chutong Meng, Haohe Liu, Qiuqiang Kong, Tom Ko, Chengqi Zhao,
  Mark~D Plumbley, Yuexian Zou, and Wenwu Wang,
\newblock ``Wavcaps: A chatgpt-assisted weakly-labelled audio captioning
  dataset for audio-language multimodal research,''
\newblock {\em arXiv preprint arXiv:2303.17395}, 2023.

\bibitem{chen2020vggsound}
Honglie Chen, Weidi Xie, Andrea Vedaldi, and Andrew Zisserman,
\newblock ``Vggsound: A large-scale audio-visual dataset,''
\newblock in {\em ICASSP 2020-2020 IEEE International Conference on Acoustics,
  Speech and Signal Processing (ICASSP)}. IEEE, 2020, pp. 721--725.

\bibitem{piczak2015esc}
Karol~J Piczak,
\newblock ``Esc: Dataset for environmental sound classification,''
\newblock in {\em Proceedings of the 23rd ACM international conference on
  Multimedia}, 2015, pp. 1015--1018.

\bibitem{gemmeke2017audio}
Jort~F Gemmeke, Daniel~PW Ellis, Dylan Freedman, Aren Jansen, Wade Lawrence,
  R~Channing Moore, Manoj Plakal, and Marvin Ritter,
\newblock ``Audio set: An ontology and human-labeled dataset for audio
  events,''
\newblock in {\em 2017 IEEE international conference on acoustics, speech and
  signal processing (ICASSP)}. IEEE, 2017, pp. 776--780.

\bibitem{song2020score}
Yang Song, Jascha Sohl-Dickstein, Diederik~P Kingma, Abhishek Kumar, Stefano
  Ermon, and Ben Poole,
\newblock ``Score-based generative modeling through stochastic differential
  equations,''
\newblock in {\em International Conference on Learning Representations}, 2020.

\bibitem{wu2023large}
Yusong Wu, Ke~Chen, Tianyu Zhang, Yuchen Hui, Taylor Berg-Kirkpatrick, and
  Shlomo Dubnov,
\newblock ``Large-scale contrastive language-audio pretraining with feature
  fusion and keyword-to-caption augmentation,''
\newblock in {\em ICASSP 2023-2023 IEEE International Conference on Acoustics,
  Speech and Signal Processing (ICASSP)}. IEEE, 2023, pp. 1--5.

\end{thebibliography}

\end{document}